\documentclass[prb,twocolumn,a4paper,floatfix]{revtex4-1}
\usepackage{latexsym}
\usepackage{graphicx}
\newcommand{\figwidth}{2.7 in}

\usepackage{subfigure}    
\usepackage{verbatim}
\usepackage{amsmath}
\usepackage{color}
\usepackage{xcolor}
\usepackage{soul}

\begin{document}
\title{The interface between Sr$_2$RuO$_4$ and Ru-metal inclusion--Implications for its superconductivity} 
\author{Soham S. Ghosh$^{(1,2)}$}
\author{Yan Xin$^{(2)}$}
\author{Zhiqiang Mao$^{(3)}$}
\author{ Efstratios Manousakis$^{(1,2,4)}$}
\affiliation{
$^{(1)}$ Department  of  Physics, 
  Florida  State  University,  Tallahassee,  Florida  32306-4350,  USA\\
$^{(2)}$ National High Magnetic Field Laboratory,
  Florida  State  University,  Tallahassee,  Florida  32306-4350,  USA\\
$^{(3)}$ Department of Physics, 
  5024 Percival Stern Hall, Tulane University, New Orleans, LA 70118 USA\\
$^{(4)}$Department   of    Physics,   National and Kapodistrian 
University    of   Athens,
  Panepistimioupolis, Zografos, 157 84 Athens, Greece
}
\date{\today}
\begin{abstract}
  Under various conditions of the growth process, when the presumably
  unconventional superconductor Sr$_2$RuO$_4$ (SRO) contains micro-inclusions of Ru metal, the
superconducting critical temperature increases significantly. 
An STEM study shows a sharp interface geometry which
allows crystals of SRO and of Ru-metal to grow side by side
by forming a commensurate superlattice structure 
at the interface.
In an attempt to shed light as to why this happens, we investigated the atomic structure
and electronic properties of  the interface between the oxide and the 
metal micro-inclusions using density functional theory (DFT) calculations. 
Our results support the observed structure indicating that it is energetically favored 
over other types of Ru-metal/SRO interfaces. We find that a
$t_{2g}$-$e_g$ orbital mixing occurs at the interface with
significantly enhanced magnetic moments. Based on our findings, 
we argue that an inclusion mediated interlayer coupling reduces
phase fluctuations of the superconducting order parameter which could explain the observed enhancement of 
the superconducting critical temperature in SRO samples containing micro-inclusions.

\end{abstract}

\maketitle

\section{Introduction}

The superconducting state of Sr$_2$RuO$_4$ (SRO)
has been intensively studied\cite{RevModPhys.75.657} since its discovery 
in 1994\cite{Maeno1994}. SRO is a layered perovskite 
oxide sharing the same structure as La$_2$CuO$_4$, one of the 
parent compounds of the cuprate superconductors,  and it is
believed to be a $p$-wave superconductor with odd spin-triplet
pairing.\cite{RevModPhys.75.657,doi:10.1143/JPSJ.81.011009,Ishida1997,Ishida1998,Nishizaki1998,Nelson2004,Jang2011}
A chiral orbital order parameter of the form $p_x+ip_y$ has been suggested
by time-reversal symmetry breaking experiments\cite{Luke1998,PhysRevLett.97.167002}.
The role of strong ferromagnetic spin fluctuations in mediating superconductivity has been pointed out\cite{Mazin1997}.
Theoretical\cite{Rice1995,baskaran1996} and experimental investigations which were carried out before
2002 have been reviewed\cite{RevModPhys.75.657}. Quantum oscillation experiments\cite{Mackenzie1996}  indicate that 
the normal state can be understood as a two dimensional Fermi liquid\cite{RevModPhys.75.657}.

The electronic properties have been studied by a number of methods which are summarized in
Ref.~\onlinecite{RevModPhys.75.657}, including 
the local density approximation (LDA) method\cite{Oguchi1995, Singh1995}
and the generalized gradient approximation (GGA)\cite{deBoer1999, Hao2014} method. Depending
on the exchange correlation functional used, GGA predicts either a nonmagnetic
state\cite{Hao2014} or a antiferromagnetic (AF) state\cite{deBoer1999} 
with ferromagnetically ordered RuO$_2$ basal planes.

In more recent studies, superconductivity 
in bulk SRO is found to be enhanced  under uniaxial $\left \langle 001 \right \rangle$\cite{Kittaka2010}, 
$\left \langle 100 \right \rangle$ and $\left \langle 110 \right \rangle$ strains where in the latter two cases
strain-driven asymmetry of the lattice is believed to cause a change in symmetry of the 
superconducting order parameter\cite{Hicks283}. T$_c$ is also found to be enhanced due to dislocations\cite{Ying2013}, 
and in a system where there is an interface
of W/Sr$_2$RuO$_4$ point contacts\cite{Wang2015}. 

The unexpected enhancement of T$_c$ from 1.5 K to almost 
3 K, when micro-sized ruthenium metal inclusions are embedded inside SRO 
in the eutectic system during crystal growth\cite{Maeno1998}, is a very
interesting and unexplained phenomenon. There is evidence\cite{Mao2001} that 
the ``$3$-K" superconducting phase is unconventional with the presence of a hysteresis loop\cite{Ando1999}.
In this ``3-K" phase,  the ruthenium micro-platelets  are not uniquely 
oriented with respect to the SRO lattice. 
 This, together with large diamagnetic shielding could suggest the existence 
 of interface superconductivity at the 
Ru-SRO interface, residing primarily in SRO\cite{Maeno1998}.
Sigrist and Monien's\cite{Sigrist2001} 
phenomenological analysis postulates a superconducting state 
with different symmetry and higher T$_{c}$ than in the bulk.

\begin{figure}[htb]

    \begin{center}
            \includegraphics[width=\figwidth]{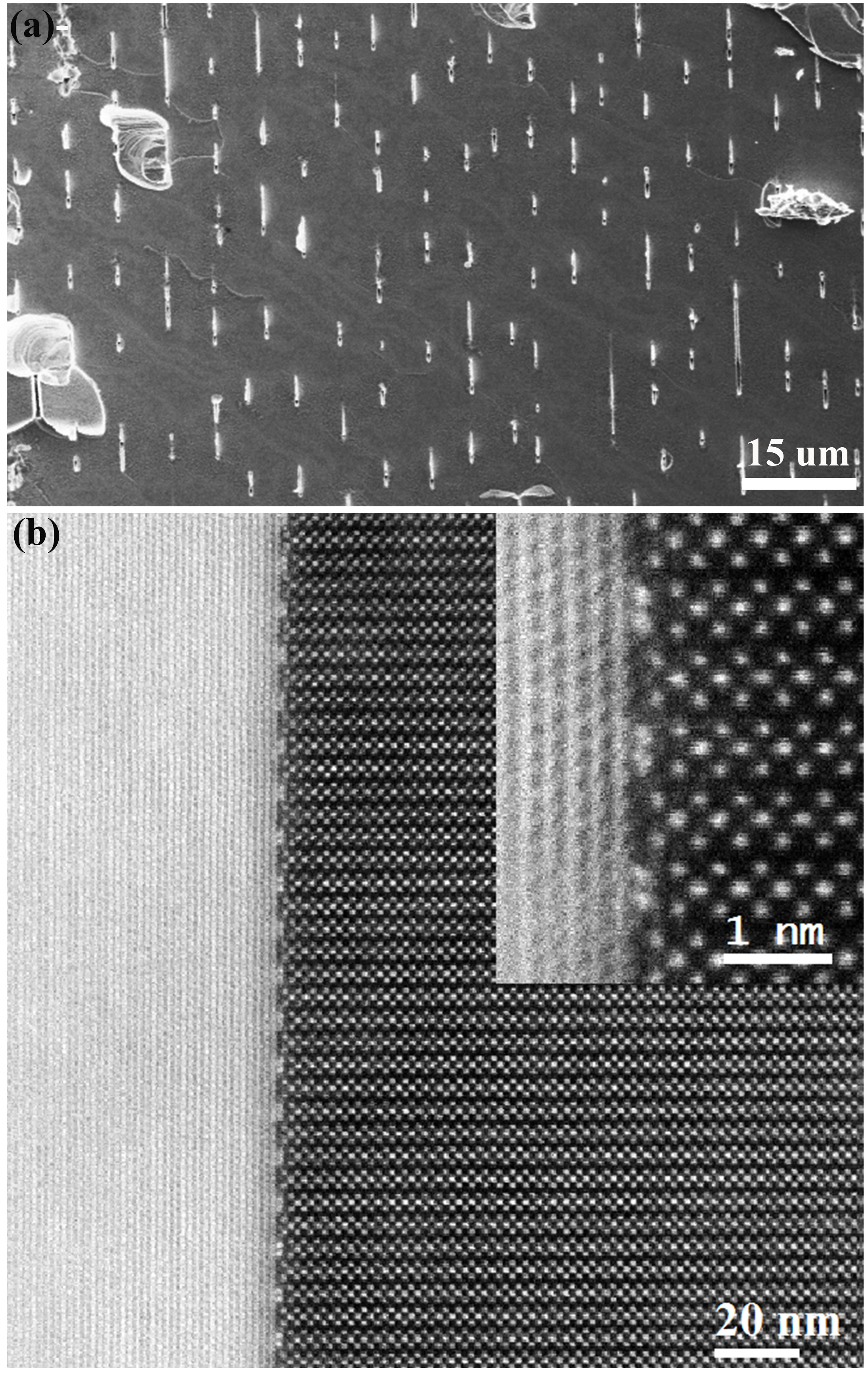}
    \end{center}
        \caption{(a) Scanning electron microscope image of a cleaved (001)
        Sr$_2$RuO$_4$ bulk crystal containing parallel Ru micro-platelets 
        (brighter contrasted short lines). 
          (b)   HAADF-STEM   image  of   the   Sr$_2$RuO$_4$/Ru
          interface  at  lower   magnification  showing  an  atomically
          straight  and  sharp  super-structured  interface,  where  the
          brighter  region on  the  left  is  the Ru metallic inclusion
          and  on  the  right  is  the Sr$_2$RuO$_4$ phase.  
          The brightest spots are Ru followed by
          Sr. O atoms  are too  faint to  be observed.   There is  no
          concentration gradient  of SRO or  Ru on either side  of the
          interface. 
          (inset) Close up view of the interface at higher
          magnification,  where   the  atomic  structure   is  clearly
          revealed. The HAADF-STEM  images were taken with  a probe of
          0.078  nm and  a  convergence  semi-angle of  21  mrad and  inner
          collection angle of 78  mrad. Brighter-contrast atoms are Ru
          atoms while Sr atoms are  less bright.
}
    \label{fig:yanfig}
\end{figure}

STEM images of a representative interface were presented in Ref.~\onlinecite{Ying2009} 
and in the present paper with higher resolution as shown in Fig.~\ref{fig:yanfig}. 
Here, we explore the atomic structure, 
stability and the electronic properties of this interface using 
density functional theory (DFT) calculations. We notice that this stable
interface is perpendicular to the SRO bulk crystallographic $a$ axis and has alternating intact meandering
Ru-O octahedra, which can be conceived as continuations of the bulk SRO RuO$_2$ planes.
Alternating pairs of Ru columns (which, as we show, are coordinated by oxygen atoms which are not visible in the STEM images) fill in the gaps created by the
meandering interface at the same periodicity as the SRO unit cell along the crystallographic $c$ axis.
These pair columns of Ru atoms have a coordination number different     
from that in the metal phase and that in the SRO phase. 
A nearly perfect hcp crystal of Ru metal grows from the next metal layer 
with a small lattice mismatch that eventually relaxes as one moves away from the interface 
and into the inclusion.

In the lowest energy interface the 
Ru-metal grows its first layer commensurate
with the SRO interface at a wavelength which nearly corresponds
to eleven Ru metal atoms for every two periods of the SRO crystal
along its own $c$ axis. Therefore, the inclusions connect RuO$_2$ planes through a commensurate 
interface which is  almost perfectly ordered at the atomic scale.

We show that this interface is stable against phase separation and it is more stable 
than other conceivable interfaces between SRO and Ru-metal. 
A spin  polarized GGA calculation yields significant magnetic
moments of the Ru atoms in the  SRO phase near the interface. Our study establishes a clear picture of 
the stable Ru-Sr$_2$RuO$_4$ interface which is important in understanding  
the unconventional ``$3$-K" phase. 

 We  also argue  that  our findings, that the Ru inclusions form a nearly atomically
perfect interface with the SRO crystal, imply the emergence of a significant 
interlayer coupling which could give rise to reduction of phase fluctuations of the
superconducting order parameter characterizing the various RuO$_2$ planes.
This is expected to lead  to an enhancement of the superconducting critical
temperature as observed in the SRO crystals with Ru inclusions.

The paper is structured as follows. In Sec.~\ref{DFT}, we present 
the computational details of our DFT calculations of the observed interface 
structure and stability consideration of various terminations. In Sec.~\ref{results}, we
analyze STEM observations based on the DFT-based calculations and also
discuss our detailed results obtained from these calculations. In Sec.~\ref{superconductivity}, 
we discuss the possible causes of increased T$_c$ based on our findings.
Last, in Sec.~\ref{conclusions}, we highlight our conclusions and the implications of this work.

\section{DFT study of the interface structure}
\label{DFT}
The platelet inclusions are the result of excess Ru
(more RuO$_2$ in the mixture than needed to make stoichiometric SRO) 
in the initial mixture. We will show by means of DFT calculations that
the interface seen in our STEM images is low enough in energy to be
preferable than macroscopic phase separation of SRO and Ru metal.
In addition, we will show that other related interfaces are energetically
higher than the one shown in Fig.~\ref{fig:yanfig}.

Sr$_2$RuO$_4$ occurs in bulk in the body-centered tetragonal
structure like the high temperature 
superconductor La$_2$CuO$_4$\cite{Maeno2001}. Ru,
the brightest atoms in the STEM images in the SRO phase, 
 and the planar O(1) atoms form a 
two-dimensional square lattice with Ru-O bond length being $1.95$ \AA\, which is less than the 
sum of the ionic radii of Ru$^{4+}$ and O$^{2-}$, suggesting planar
hybridization\cite{Singh1995}.  
The apical O(2) height above and below is larger at $2.06$ \AA.

\subsection{Analysis of the observed interface}
First, we notice in Fig.~\ref{fig:yanfig} and in Fig.~1 of Ref.~\onlinecite{Ying2009} that the interface is
perpendicular to the bulk [001] SRO direction and is terminated at alternating intact
meandering Ru-O octahedra, which are continuations of the bulk SRO RuO$_2$
planes, and alternating pairs of Ru columns filling in the gaps created by the meandering interface. 
Notice that, as seen in Fig.~\ref{fig:yanfig},
the Ru ions of the metallic interface, without any significant change of the unit-cell size,
form an interface at a commensurate wavelength 
nearly
twice that of the SRO unit-cell size in the [001] direction (corresponding to eleven unit cells of pure  Ru crystal).
The commensurate growth leaves a small lattice mismatch between the SRO layers near the interface and those 
deeper in the bulk. This causes a strain that grows with system size until it becomes energetically favorable to 
 produce dislocations\cite{Ying2013} relieving the strain. As seen in the transmission electron microscopy (TEM) images 
 in Ref.~\onlinecite{Ying2013}, the dislocations are complex structures which occur, for a sharp 
 and flat interface, over a length scale longer than that visible in Fig.~\ref{fig:yanfig}. These structures  
 are beyond the scope of our present \textit{ab initio} calculations and we focus here on the microscopic 
 length scale at which the interface is free of dislocations.
 
STEM image provides great information on the interface structure
but leaves an ambiguity on any oxygen atom near the interface and  the
Ru atom positions in the SRO [010] direction which is perpendicular to the
STEM field of view. This is the first item to address by
means of our DFT calculations, the technical details of which are discussed later. 
By examining various possibilities, we find that in their optimum positions, 
the Ru pair columns bridge diagonal oxygen atoms of the terminating SRO layer. 
On one side of the interface these Ru pair columns form a similar type bond 
to oxygen atoms as the 1.95 \AA\, Ru-O bond in the rutile phase of RuO$_2$,
and on the other side they have an hcp Ru metal environment.
This  arrangement leaves the meandering SRO termination layer unchanged subject to ionic relaxations, and 
is consistent with the experimental image in Fig.~\ref{fig:yanfig}.
Therefore, we believe that this part of our DFT study complements the
STEM image and we now have a complete picture of the structure of the
interface.

\subsection{Stability of the interface}

\begin{figure}[htp]
\vskip 0.2 in
\begin{center}
\includegraphics[width=\figwidth]{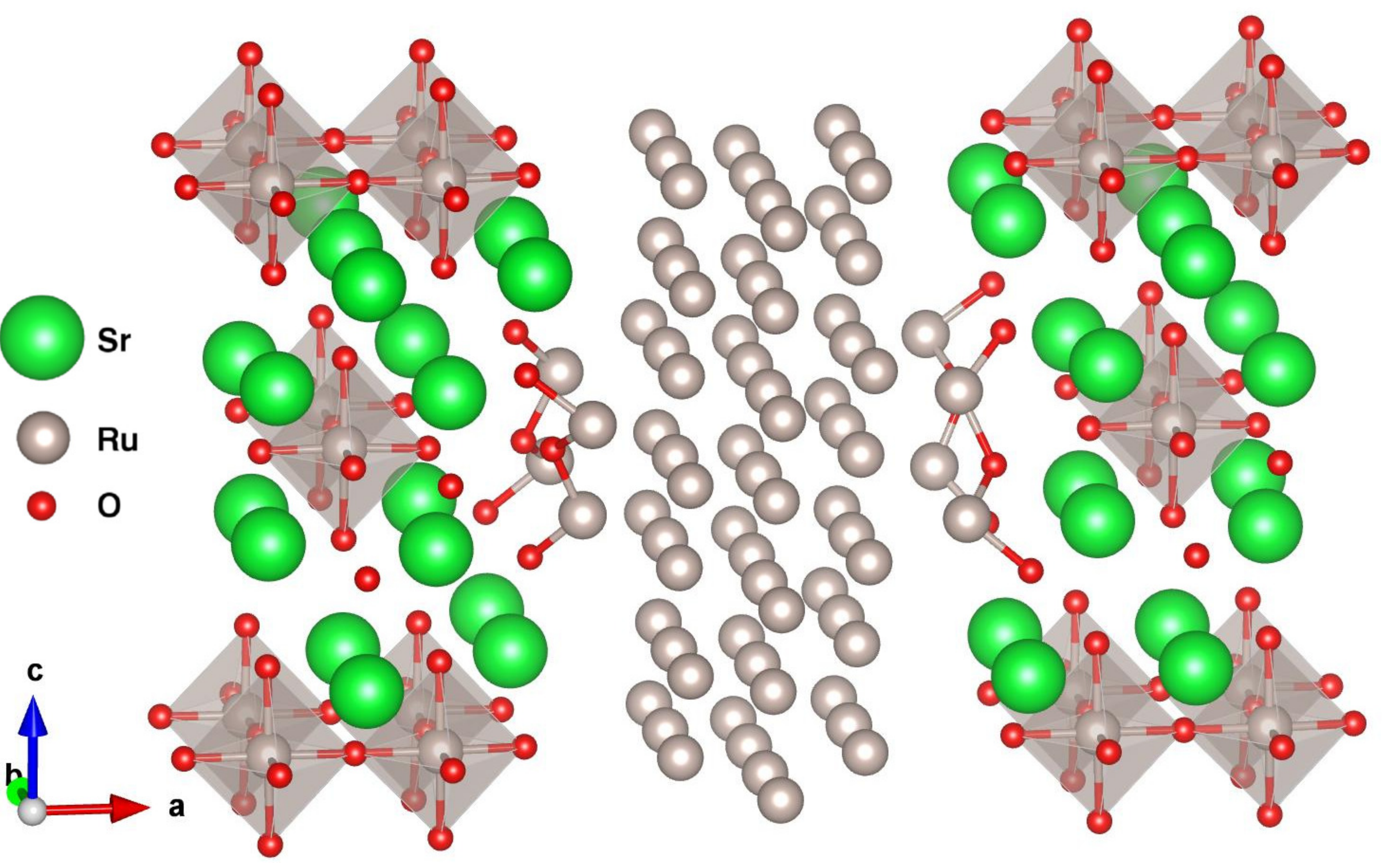}
\end{center}
\caption{(color-online) 
         Sr$_2$RuO$_4$-Ru supercell used for spin-GGA calculations of the 
         heterojunction as generated by VESTA software package. 
         There are three different Ru-O bond lengths: The Ru-O bond lengths
         in the interface columns are the shortest, followed by the in-plane Ru-O(1) distances, 
         followed by the Ru-O(2) bond lengths. The interface Ru atoms columns mediate between 
         the SRO phase and the Ru metallic phase. We show part of the next repeated image for clarity. 
         SRO lattice vectors are shown.}   
\label{fig:geometry}
\vskip 0.2 in
\end{figure}

\begin{figure}[htp]
\vskip 0.2 in
\begin{center}
\includegraphics[width=\figwidth]{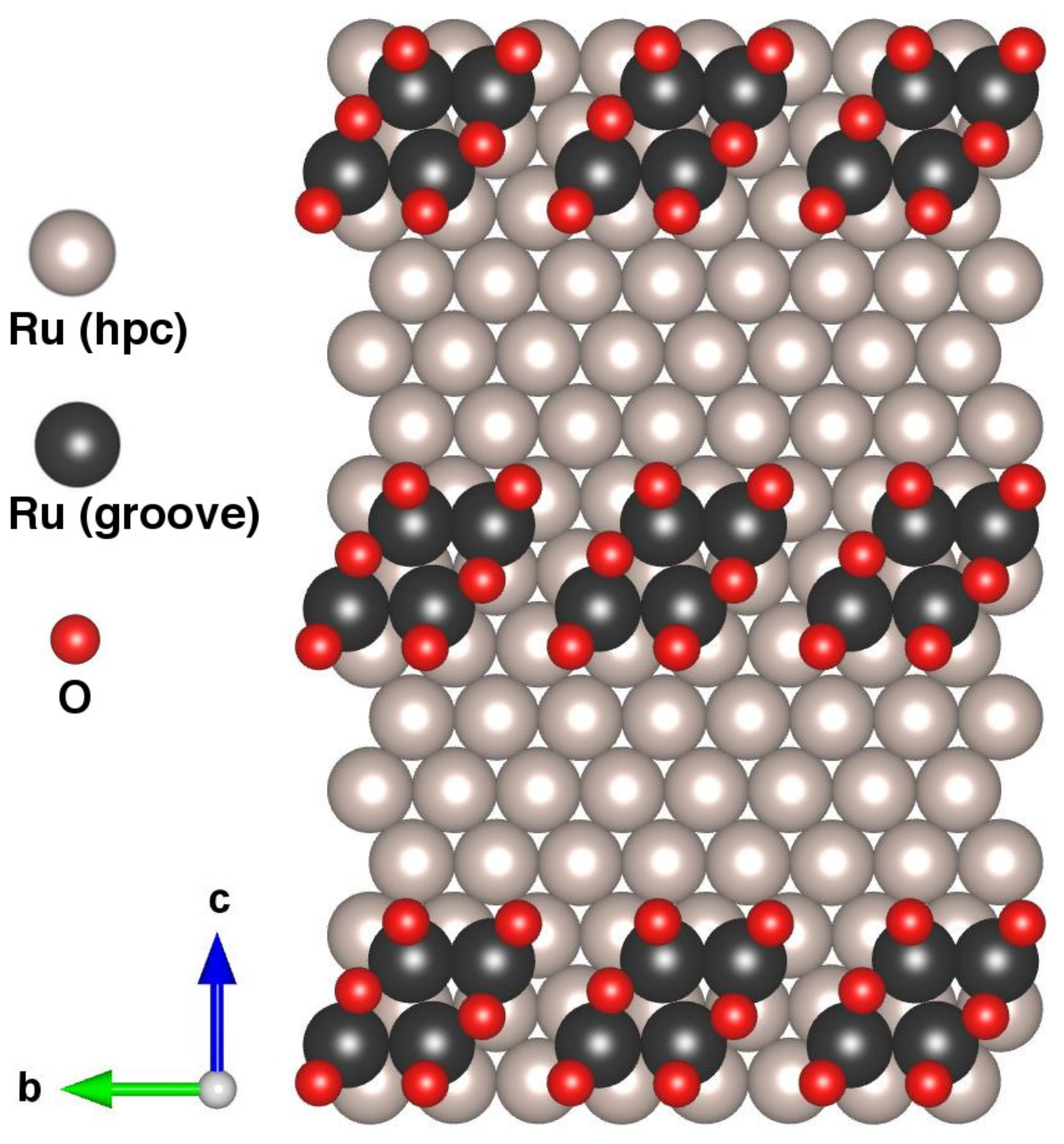}
\end{center}
\caption{(color-online)
                   The SRO $b-c$ plane showing the two layers of the SRO-Ru interface, 
                   formed by the interface Ru pair columns and the
                   terminating layer of the hcp metal. The unit cell lengths are slightly 
                   different from ideal bulk values (with details in the text). The triangular 
                   lattice of the Ru metal layer and the rectangular lattice
                   of the interface Ru pairs are commensurate with $7.81$ \AA\ $\times 13.52$ \AA\, wavelengths 
                   in the SRO [010] and [001] directions respectively. The Ru columns are 
                   situated in the low energy valleys of the triangular lattice.
                   The experimental wavelength is at least twice in each direction with an increased 
                   fraction of misalignment where some interface columns are located away from the hcp valleys, but the 
                   idea commensuration between a rectangular and a triangular lattice is preserved.}

\label{fig:interface_grooved}
\vskip 0.2 in
\end{figure}

\begin{figure}[htp]
\vskip 0.2 in
\begin{center}
\includegraphics[width=\figwidth]{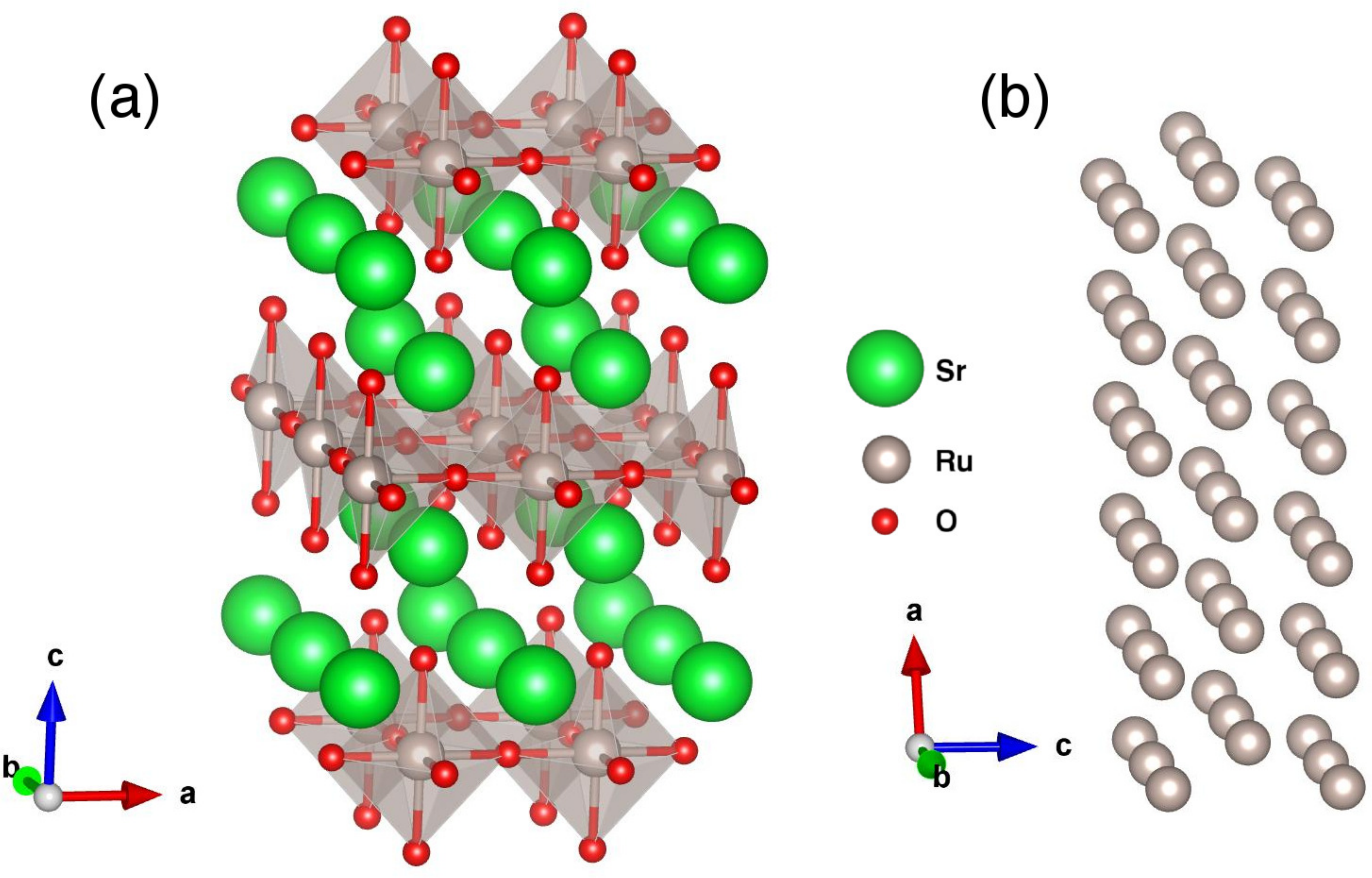}
\end{center}
\caption{(color-online) 
         (a) The $2.5$-layer thick Sr$_2$RuO$_4$ slab with equivalent (010) surfaces used to compute the energy 
         of (001) linear strain changing the $c$-lattice vector length from $12.9$ \AA\, to $13.52$ \AA\, while keeping the 
         $a$ and $b$-vectors same as in bulk. (b) The $1.5$-layer thick ruthenium metal slabs in its hcp structure with 
         (001) surfaces used to compute the energy of planar compression. The $a$, $b$ lattice vectors were each reduced from $2.73$ \AA\, to $2.61$ \AA\,. In both calculations
         we used at least $15$ \AA\, of vacuum between repeated images.}
\label{fig:slabs}
\vskip 0.2 in
\end{figure}

\begin{figure}[htp]
\vskip 0.2 in
\begin{center}
\includegraphics[width=\figwidth]{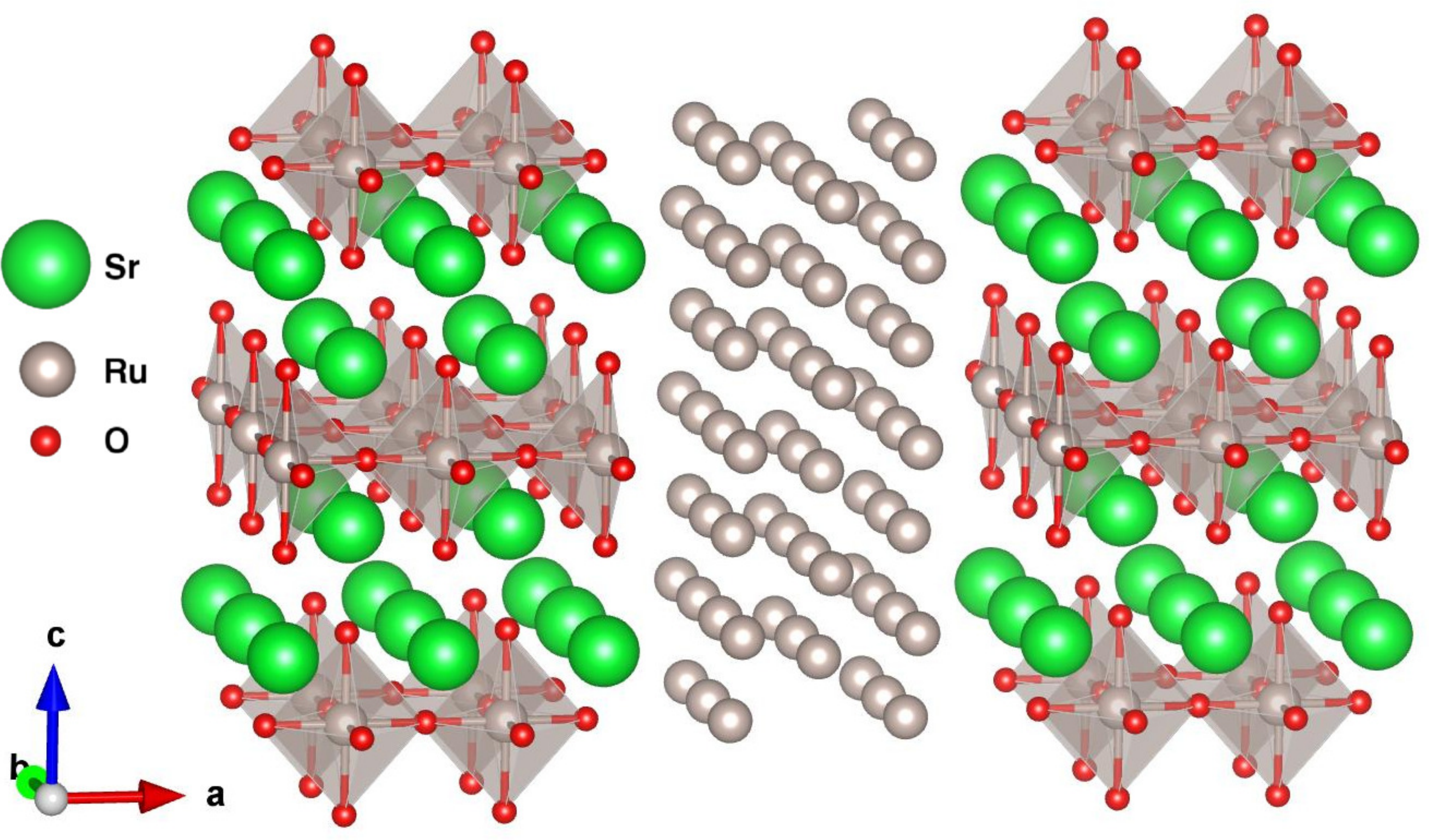}
\end{center}
\caption{(color-online)
         A hypothetical interface between SRO [100] and Ru hcp [001] surfaces, created from the slabs in Fig.~\ref{fig:slabs} by
         bringing them together gradually and allowing the atoms to relax till the optimum interlayer distance at the
         interface is reached. This geometry is similar to that in Fig.~\ref{fig:geometry} except in the absence
         of the Ru interface columns. The hcp c-axis is parallel to the SRO [100] direction. Part of the next repeated image
         is shown for clarity.}
\label{fig:flat_geometry}
\vskip 0.2 in
\end{figure}

\begin{figure}[htp]
\vskip 0.2 in
\begin{center}
\includegraphics[width=\figwidth]{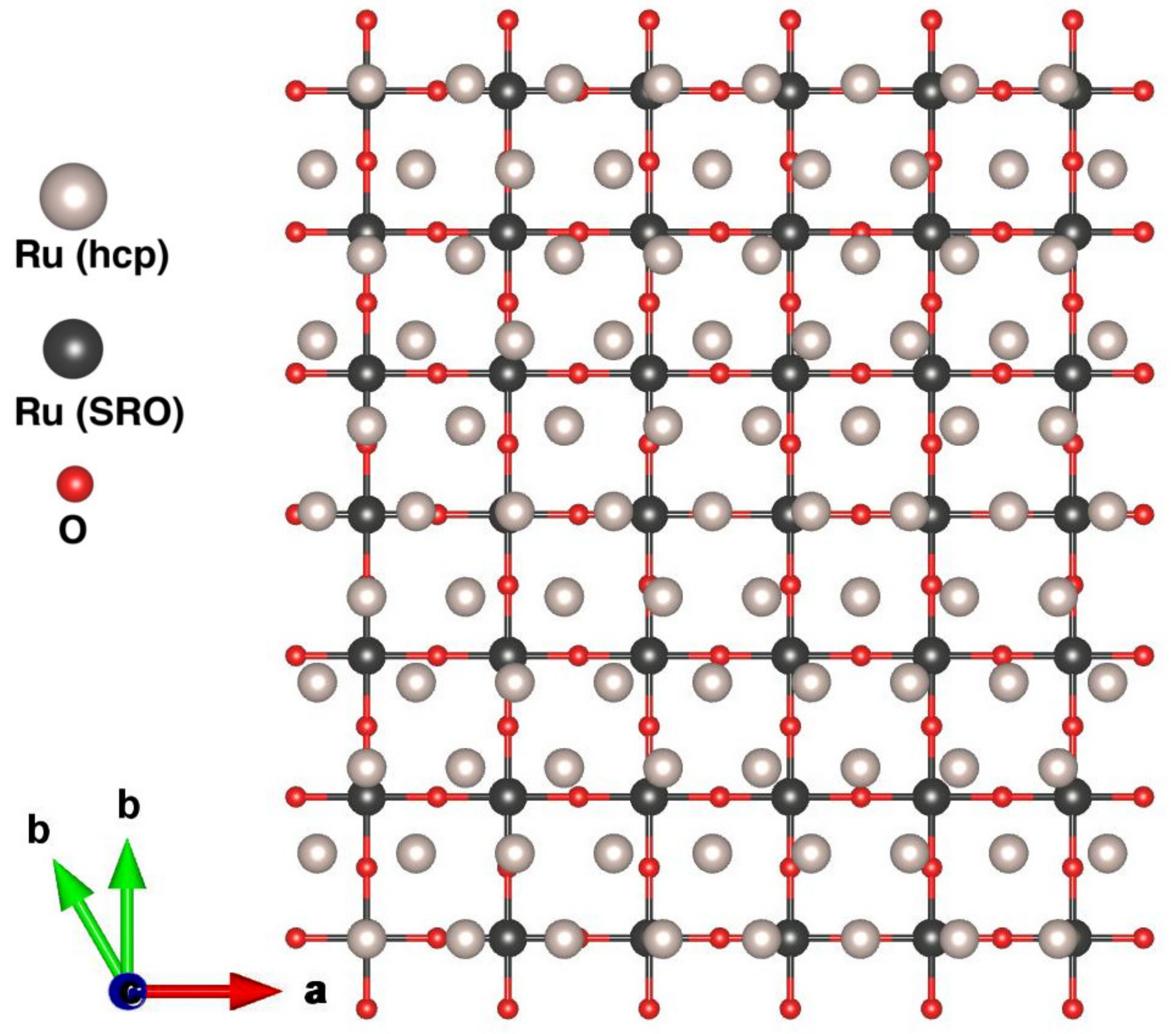}
\end{center}
\caption{(color-online)
         A possible interface between (001) SRO and hcp Ru metal with a $19.52$ \AA\, $\times 23.42$ \AA\, superlattice
         where after a $0.8$ \% strain on the hcp plane, five planar unit cells of  RuO$_2$ is commensurate with seven unit cells 
         of the Ru metal in the SRO [100] direction
         and six unit cells of  RuO$_2$ is commensurate with five unit cells of the Ru metal in the SRO [010] direction. Most of the
         atoms at the interface are frustrated and experience the laterally averaged potential leading to very little 
         energy gain. Both SRO and Ru-metal lattice vectors are shown.}
\label{fig:overlay001}
\vskip 0.2 in
\end{figure}

The implementation of the observed structure by DFT calculation
ideally requires at least eleven bulk unit cells of 
the hcp Ru metal phase commensurate with two conventional bulk unit cells of the SRO along the SRO [001] axis, and at least six 
bulk unit cells of the hcp Ru commensurate with four bulk units cells of the SRO along the SRO [010] direction. 
Such a supercell proved computationally unfeasible, particularly due to the large number of ruthenium atoms. 
Instead, to  keep the supercell size reasonable, 
we used a supercell geometry with a stretched unit cell length of $13.52$ \AA\, along the SRO [001] direction 
and double the SRO bulk unit cell length in the SRO [010] direction.

In this reduced supercell containing 128 atoms, shown in 
Fig.~\ref{fig:geometry}, one conventional bulk SRO unit cell is commensurate with six bulk hcp Ru unit cells 
along SRO [001] and two bulk SRO unit cells are commensurate with three bulk metal unit cells in 
the SRO [010] direction. The atoms have been relaxed to their final positions. 
This geometry stretches the SRO c-axis by 5 \% and compresses the a, b axes of the hcp Ru by 
4 \% each. The apical O(2) height is increased from $2.06$ \AA\, to $2.09  $ \AA.\
In Fig.~\ref{fig:interface_grooved} we show the SRO $b-c$ plane of the two layers of the SRO-Ru interface formed by the interface Ru pair columns and the terminating layer of the hcp metal.
Eventhough we will use this smaller super-cell in our DFT implementation, by carrying out various types of optimization calculations
it will become reasonably clear that the observed structure is indeed the most energetically favorable among various other plausible atomic configurations.

First, we find that the energy needed for breaking apart the interface of our 
reduced structure to create the two constituent slabs - SRO with [100] surfaces and Ru metal with 
[001] surfaces (Fig.~\ref{fig:slabs}) - to be $13.16$ eV, a significant amount of energy. 

Second, we compare the energy of our reduced structure to the energy of a similar supercell but with flat 
SRO [100] - Ru [001] interface (shown in Fig.~\ref{fig:flat_geometry}) which we compute by bringing 
the SRO and Ru phases closer in small steps and relaxing the ions around their positions to find 
the minimum of the energy. We find that this supercell is higher 
in energy by $8.37$ eV compared to our reduced meandering geometry in Fig.~\ref{fig:geometry}. 
Thus the meandering octahedra and the interface ruthenium columns are necessary to  
stabilize the interface. In both the above calculations, there are 128 atoms on either side of the equation and we account for the 
balance of atoms in the non-meandering structures by putting the extra Ru pair columns of 
the interface in a ruthenium bulk phase, which is the most stable ruthenium phase. 
To describe why the meandering Ru interface columns are necessary, consider the two interface 
layers ($b-c$ plane of SRO) in Fig.~\ref{fig:interface_grooved}. 
We note that the interface Ru columns form a rectangular planar lattice (which is commensurate 
with the same periodicity of our reduced interface geometry) adjacent to the 
triangular lattice of the terminating Ru metal plane. Furthermore, 
the interface Ru pairs to a large extent are situated in potential valleys where the atoms of the next Ru hcp layer 
would have been in absence of an interface.

Third, having established the necessity of the meanders, we argue that the larger
interface with eleven unit cells of Ru-metal in the SRO [001]  direction and six Ru-metal unit cells in the SRO [010] direction 
is more stable than our computed reduced interface as follows.

The reduced geometry suffers from
artificial planar compression of its Ru metal phase and uniaxial strain of its SRO phase but it accommodates the
interface Ru pair columns in the grooves created by the terminating hcp Ru layer.
By sliding the Ru metal phase across the interface in Fig.~\ref{fig:geometry}, we find that this is indeed an
energy minimum. The larger experimentally observed periodicity breaks this symmetry and at least some of the Ru
column pairs are misaligned with respect to the hcp layer. This leads to frustration in the terminating hcp layer of the metal, as 
can be observed directly in Fig.~\ref{fig:yanfig}.
To a first approximation, the price of lattice length manipulation can be calculated by computing 
the sum of the individual energy losses caused by
separately compressing an ideal  Ru metal slab and stretching an ideal SRO slab (Fig.~\ref{fig:slabs}) to their
respective values necessary to create the reduced supercell shown in Fig.~\ref{fig:geometry}. We find this energy
to be $8.54$ eV. On the other hand, the energy loss due to non alignment of the
Ru interface pair columns in the experimentally observed geometry can be upper bounded by sliding the Ru metal layers of the reduced geometry along the interface, thus moving the Ru pair columns away from the 
potential valleys of the hcp layer, until we reach an energy maximum. 

The maximum cost of misalignment of the interface Ru pair columns is $4.49$ eV, less than the cost of stretching and compressing the
constituent phases. It follows that in any interface between the SRO (100) surface and Ru (001) surface,
 meandering Ru interface columns with the  experimentally observed periodicity is the most stable structure.

 As a final part of our stability argument, we consider the possibility that there can be other
types of interfaces without these interface Ru columns which are more stable that ours.
In particular, the (001) surface of SRO can be cleaved with a terminating
SrO or RuO$_2$ layer and its interface with a ruthenium hcp layer can be conceived. We show in Fig.~\ref{fig:overlay001}
a commensurate geometry at the interface, in which a $5 \times 6$ RuO$_2$ superstructure (defined along 
the SRO a and b-axes respectively) is commensurate
with a $7 \times 5$ superstructure of the metal, after accounting for a $0.8$ \% uniaxial strain on the
Ru metal. Most of the atoms of each phase in this superlattice are randomly oriented with those of the  other phase and therefore
each layer at the interface will experience the laterally averaged potential of the other layer.
This is arguably a small energy gain of the order of a few tens of meV for each SRO formula 
unit. In fact, any interface without the periodic meanders observed in our experiment is likely to be disfavored for the same reason.
A straightforward GGA computation shows that our reduced structure is lower in energy
by $1.72$ eV compared to the sum of energies of a Ru metal slab similar to that shown in Fig.~\ref{fig:slabs}(b) and
the appropriate amount of bulk SRO, a surprising result. The  experimentally observed superlattice should be
 even more stable. 
We conclude therefore that the observed meandering interface is favored even over phase separation of the 
eutectic mixture between bulk SRO and Ru. Therefore, we cannot think of 
any other interface which can compete with the one observed by our STEM study. 

Lastly, the following question arises. Since we find that the observed interface lowers the energy 
with respect to bulk SRO and a semi-infinite Ru metal, why the system
does not try to create more such interfaces and instead it grows mesoscopic size inclusions.
Indeed, our findings indicate that the lowest energy state of such a system of SRO with 
excess Ru metal should be a state with a high density of such interfaces separated by a 
microscopic-size length. However, the combined system was created under non-equilibrium conditions 
of the eutectic mixture which do not allow for the system to search for a global lowest energy 
state. First, at a relatively short-time scale the free energy is only locally minimized and, then, the 
system freezes in a macroscopic state of domains which require an overwhelmingly large amount of time 
to find the state which is the global minimum. More specifically, once a few layers of Ru metal 
have grown, it becomes locally energetically favorable for more Ru atoms from the excess of Ru, to 
attach themselves to those existing Ru metal layer. Forming another interface which combines a 
simultaneous and coherent arrangement of many atoms is a much slower process (i.e., a low entropy state) than 
simply adding to the existing Ru metal layer an additional single Ru 
atom. This means that the path to the actual ground state is ``narrow" and requires a very slow process and, 
as a result, the system gets stuck in other metastable local free-energy minima. Thus, although a high 
density of such interfaces is preferred by taking into consideration just the energy of the system, 
the meandering termination layer of SRO with intact RuO$_2$ octahedra and interface Ru columns 
are long-range phenomena generally suppressed by the large entropy present in the high temperature eutectic mixture.

\subsection{Computational details}

Spin-GGA computations were performed using plane-wave basis set (cutoff of 540 eV) 
with the projected augmented wave methodology\cite{Blochl} used to describe the wavefunctions of the 
electrons as implemented in the \textsc{VASP} package\cite{Shishkin3,Fuchs,Shishkin2,Shishkin1}, 
using the Perdew-Burke-Ernzerhof (PBE) exchange correlation functional\cite{PBE}.
 The 4\textit{s}, 4\textit{p}, 5\textit{s} electrons of strontium, 
the 5\textit{s}, 4\textit{d}, 4\textit{p}, 4\textit{s} electrons of Ru
and the  2\textit{s}, 2\textit{p} electrons of the oxygen
were treated as valence electrons. The Brillouin zone of the 128-atoms supercell 
 with the meandering interface geometry (Fig.~\ref{fig:geometry})
and the 124-atoms supercell with flat interfaces (Fig.~\ref{fig:flat_geometry})  were sampled with $ 1 \times 6 \times 4$
\textit{k}-point grid, and 60 \textit{k}-points were
used to compute the electronic density of states (DOS). Increasing the \textit{k}-point grid from 
$1\times 6 \times 4$ to $1 \times 7\times 5$ leads to a negligibly small change of 
the total energy of the meandering interface geometry by 0.04 eV  and an increase of the 
moments of the magnetic ruthenium atoms by 0.08 $\mu_{B}$. 
The $70$-atoms SRO slab (Fig.~\ref{fig:slabs}(a)) is $2.5$ layers 
thick with symmetric $2 \times 1$ $b-c$ surfaces. We have used  
a tetragonal geometry with $b = 3.90$ \AA\, and $c = 12.90$ \AA\, and sampled the Brillouin zone 
with a $ 1 \times 8 \times 4$ \textit{k}-point grid.
For the 54 atoms Ru metal slab (Fig.~\ref{fig:slabs}(b)) which is $1.5$ layers thick with 
 a $6 \times 3\times 1.5$ structure, we have used an hcp unit cell length of $2.739$ \AA\, and sampled the 
Brillouin zone with $ 4 \times 8 \times 1$ \textit{k}-point grid. For both the slabs, 
we have used a vacuum layer at least $15$ \AA\, thick.
All the supercells was structurally relaxed while keeping the cell shape and cell volume 
fixed until the forces were converged to less than 10 meV/\AA\, for each ion.\\

\section{Electronic properties}
\label{results}

\begin{figure}[htp]
\vskip 0.2 in
\begin{center}
\includegraphics[width=\figwidth]{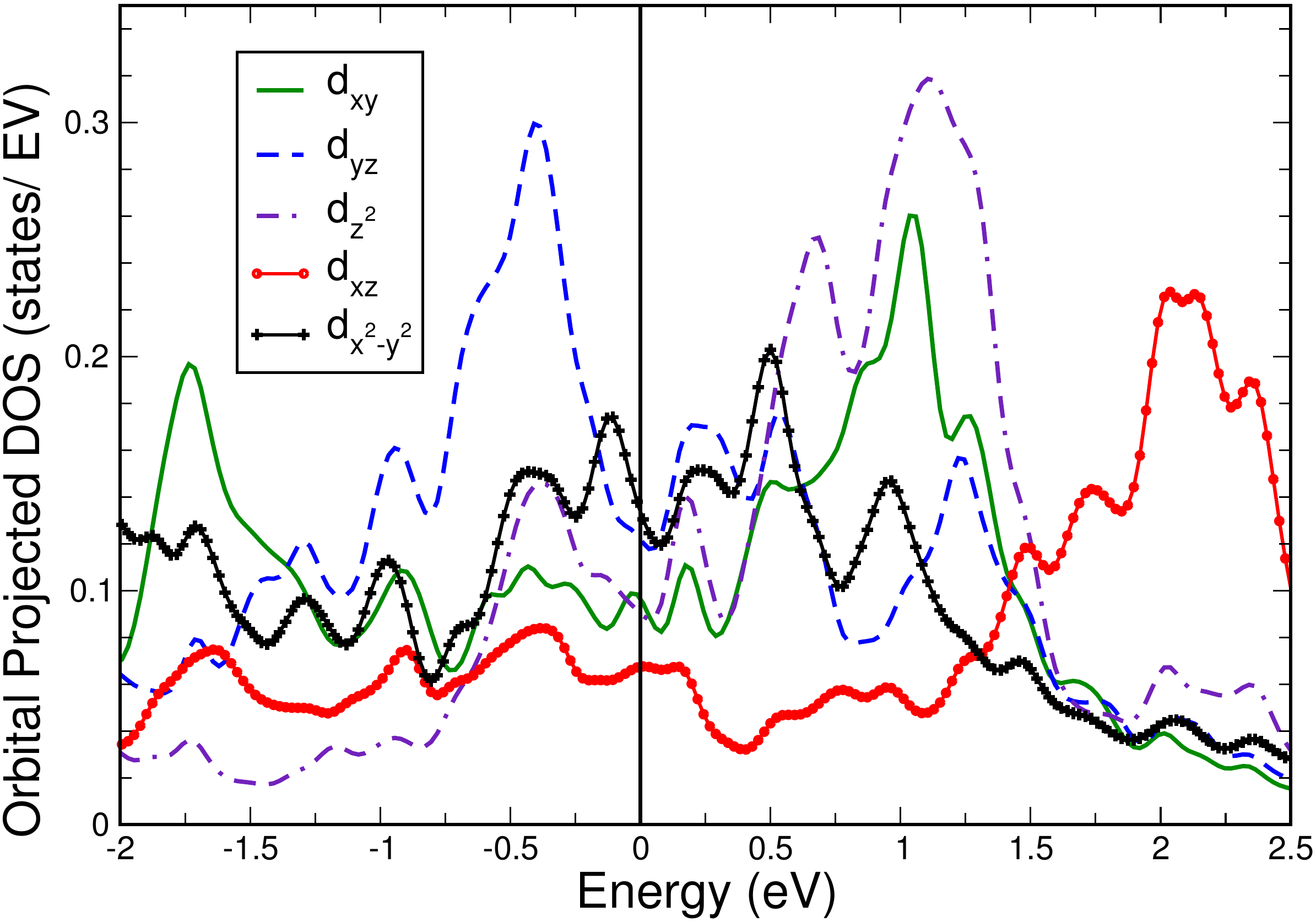}
\end{center}
\caption{
         (color-online) $Y_{lm}$-projected density of states for a Ru atom from the interface 
         column pair. The atom has a O$_2$ environment like in the rutile RuO$_2$ phase with bond angle of 
         115$^{\circ}$ and Ru-O bond length of 1.83 \AA. On the other hand, it also neighbors the Ru metal closed-pack 
         layer at the interface. Both these environments influence its electronic structure.
} 
\label{fig:Ru_8dos}
\vskip 0.2 in
\end{figure}

Our computed geometry (Fig.~\ref{fig:geometry}) shows a 
1.8$^{\circ}$ rotation of the RuO$_6$ octahedron on the RuO$_2$ planes along 
with small amounts of buckling. We find no magnetic moments 
in the interface columns of Ru atoms and metallic phase Ru atoms, but strong magnetic 
moments in the Ru atoms ($M_{Ru} = 1.532\, \mu_{B}$)  in the SRO phase. 
GGA calculations have previously predicted\cite{Matzdorf2000} surface ferromagnetism in SRO where it was 
stabilized by a large (9$^{\circ}$) surface octahedra rotation and consequent band narrowing. To ensure that 
this is not  purely an effect of the (001) strain, we have performed spin-GGA calculations 
of bulk SRO with stretched c-axis values. The octahedra rotations are absent and we find 
a ferromagnetic ground state but with much smaller magnetic moments ($M_{Ru} = 0.220\, \mu_{B}$).

In the interface Ru atom columns, there are two different types of RuO$_2$ bonds in each pair. One has bond angle 
close to 90$^{\circ}$ and Ru-O bond length close to the planar Ru-O(1) bond length, whereas the other has bond angle 
$\sim$ 115$^{\circ}$ and Ru-O bond length $\sim1.83$ \AA. Both Ru atoms lack the planar square lattice 
coordination and are expected to have different orbital structure compared to those in SRO. Indeed, we find 
significant $t_{2g}$-$e_{g}$ mixing in each of them. Figure~\ref{fig:Ru_8dos} shows the density of states of the Ru 
atom which has a RuO$_2$ bond angle of $\sim$ 115$^{\circ}$. Both the $d_{x^{2}-y^{2}}$ and $d_{z^{2}}$ states 
are pulled down below the Fermi level and mixed with $t_{2g}$ orbitals. $t_{2g}-e_{g}$ mixing was found at 
 the well-studied SrTiO$_3$/LaAlO$_3$ interface\cite{Pavlenko2012}, where an $e_{g}$ splitting was caused by 
oxygen vacancy and gave rise to magnetic order. Here, it is caused by severe Ru-O hybridization and non planar 
RuO$_2$ geometry. 

\section{Implications for superconductivity}
\label{superconductivity}

The observation that the ``3-K" superconductivity is unconventional with the 
presence of a hysteresis loop\cite{Ando1999}, together with our findings 
could point to a possible coexistence of short-range ferromagnetic order and spin-triplet
superconductivity\cite{Dikin2011,Li2011,Saxena2000}, or presence of two dimensional unconventional
superconductivity close to a ferromagnetic instability. In addition, an enhancement 
of the magnetic susceptibility on the interface of these inclusions with  
SRO should lead to an enhancement in the spin-triplet channel for pairing due to increased electron-paramagnon
ferromagnetic spin fluctuations\cite{Mazin1997} coupling.
While the spin-triplet superconducting order parameter might compete with
the ferromagnetic order parameter, the p-wave pairing order parameter, which
breaks time-reversal symmetry, could benefit from approaching closer to the ferromagnetic instability.  We find that the volume of 
these micro-inclusions is of the order of $\sim 10\, \mu m^3$ and
their surface is $\sim 50\, \mu m^2$. Therefore, the volume to surface ratio is
about $2000$ \AA\, which is only about 200 times larger than the
unit cell of the SRO along the [001] direction. If we consider
the interface magnetic moments as magnetic impurities, the system is 
in the very dilute limit (of the order of 0.01 \%), 
and it might seem hard to understand that these magnetic impurities would give rise to
such a large effect on the superconducting transition temperature. 
It is already known\cite{Ortmann2013} that approximately 0.7 \% Co magnetic
impurities in pure Sr$_2$RuO$_4$ lead to a long lived ferromagnetic state.
Furthermore, the superconducting T$_C$ depends exponentially
on the electron-paramagnon coupling, i.e., $k_B T_c \sim \hbar \omega_p \exp(-m^*/\lambda)$ where  $\omega_p$ is the 
paramagnon frequency, $m^*$ is the electron effective mass enhancement and $\lambda$ is a dimensionless measure of
the electron-paramagnon coupling\cite{Mazin1997}. 
Thus, T$_c$ is expected to be very sensitive to small changes in the electron-paramagnon 
coupling caused by enhanced ferromagnetic moments found by our DFT calculations at the interface with the Ru inclusions.
However, the extremely low density of the moments makes it difficult to imagine that this is the primary reason for 
enhancement of T$_c$. 
We find the following idea more appealing as a possible explanation for the
enhancement of T$_c$ in these Ru-inclusion rich SRO crystals. 

Superconductivity is believed to arise from pairing within the RuO$_2$ layers. The
interlayer coupling between RuO$_2$ planes is very weak which is expected to lead to large amplitude phase fluctuations of the order parameter as is the case of the Cuprate superconductivity\cite{carlson}.  These phase fluctuations should lower the value of the superconducting transition temperature. 

In the case where inclusions are present,  a remarkably ordered interface geometry between  tetragonal unconventional 
superconductor Sr$_2$RuO$_4$ and hexagonal closed pack metal Ru has been discovered as illustrated in both our STEM images of Fig.~\ref{fig:yanfig} and 
justified by means of our DFT calculations (Fig.~\ref{fig:geometry}) which also reveal the structure along the perpendicular direction as illustrated in our derived highly ordered structure of Fig.~\ref{fig:interface_grooved} (a direction which is hidden from any STEM study). These interfaces clearly should lead to an effective interlayer Josephson junction coupling of the superconducting order parameter which should reduce these phase fluctuations over. This coupling produced by these ordered inclusions should lead to an enhanced T$_c$ as observed in the ``3-K" phase. We believe that the reason for the enhancement of T$_c$ due to the inclusions is different from 
the reason that causes the strain driven increase of T$_c$ in pure bulk SRO 
as seen in Ref.~\onlinecite{Hicks283}. In the latter  case  the increase is due to the fact that the strain
affects the symmetry character (i.e., $p_x + i p_y$) of the superconducting order parameter as argued in 
Ref.~\onlinecite{Hicks283}. 

It has been shown experimentally\cite{Maeno1998} that inclusions increase the interlayer coherence length 
and significantly reduce the anisotropy of superconductivity: $\xi_{ab}(0)/\xi_{c}(0) = 3.6$
in the ``3-K" phase with metals inclusions as compared to 20 for the 1.5 K phase SRO.

\section{Conclusions and implications}
\label{conclusions}

A remarkably ordered interface geometry between  tetragonal unconventional 
superconductor Sr$_2$RuO$_4$ and hexagonal closed pack metal Ru formed as metallic inclusions
during the growth process of the superconductor has been revealed by STEM studies and has been investigated and understood
in the present paper using DFT. The heterojunction is characterized by regular columns of 
Ru pairs in the SRO [001]
 direction and clean octahedra terminations of the ruthenate oxide. Using 
DFT, we have correctly reproduced the experimental structure including the positions of interface oxygen atoms
and along directions hidden to any STEM study and investigated the electronic structure of the interface. We have found rotated octahedra, 
modified Ru $d$-orbitals and enhanced magnetic moments near 
the interface in the SRO phase. Application of GGA to magnetism 
should be taken with caution since it does not correctly account for correlations, but given 
the proximity to Stoner instability, it is possible that the interface is in or energetically very close to a ferromagnetic ground state. 

Our study provides a possible explanation of the enhancement of the superconducting 
T$_C$ when Ru metal inclusions are present.
We find that these inclusions form microscopically well-ordered interfaces and structure. The interfaces acts as ``ladder" which couple the 
superconducting order parameter of a large number of RuO$_2$ SRO layers over a micrometer-size length.
These inclusions should lead to an interlayer coupling which can significantly reduce  the superconducting order parameter phase fluctuations, thereby increasing the superconducting critical temperature.

This observation opens up exciting prospects when a similar growth process is applied to the case of the cuprate superconductors. If these inclusions introduce an interlayer Josephson-Junction type 
coupling, we should expect a significant enhancement  of the superconductivity critical temperature.

\vskip 0.2 in
\section{Acknowledgments}
This work was supported in part by the U.S. National High Magnetic Field
Laboratory, which is partially funded by the NSF DMR-1157490 and 
the State of Florida. 

\appendix

\end{document}